\documentclass[prd,twocolumn,superscriptaddress,amsfonts,amssymb,amsmath,showpacs]{revtex4-2}
\usepackage{bm}
\usepackage{amsfonts}
\usepackage{latexsym}
\usepackage[latin1]{inputenc}
\usepackage{graphicx}
\usepackage{amsmath}
\usepackage{palatino}
\usepackage{mathpazo}
\usepackage{textcomp}
\usepackage{float}
\usepackage{booktabs}
\usepackage{dcolumn}
\usepackage{booktabs}
\usepackage{multirow}
\usepackage{hyperref}
\hypersetup{
    colorlinks=true,
    linkcolor=purple,
    filecolor=magenta,      
    citecolor=blue
}
\usepackage{amsmath}
\usepackage{xcolor}
\usepackage{orcidlink}
\usepackage[caption=false]{subfig}
\usepackage{commath}
\makeatletter
\long\def\dddddot#1{%
  {\mathop {#1}\limits ^{\vbox to-1.4\ex@ {\kern -\tw@ \ex@ \hbox {\normalfont .....}\vss }}}%
}
\long\def\multidots#1#2{%
  \count@=0
  {{\mathop {#2}\limits ^{\vbox to-1.4\ex@ {\kern -\tw@ \ex@ \hbox {\normalfont %
  \loop%
  \ifnum#1>\count@%
  .%
  \advance\count@ by1%
  \repeat%
  }\vss }}}}%
}
\makeatother

\begin{document}

\color{black}       

\title{Phantom cosmological model with observational constraints in $f(Q)$ gravity}

\author{S.A. Narawade\orcidlink{0000-0002-8739-7412}}
\email{shubhamn2616@gmail.com}
\affiliation{Department of Mathematics, Birla Institute of Technology and
Science-Pilani,\\ Hyderabad Campus, Hyderabad-500078, India.}
\author{B. Mishra\orcidlink{0000-0001-5527-3565}}
\email{bivu@hyderabad.bits-pilani.ac.in}
\affiliation{Department of Mathematics, Birla Institute of Technology and
Science-Pilani,\\ Hyderabad Campus, Hyderabad-500078, India.}

\date{\today}

\begin{abstract}
\textbf{Abstract}: In this paper, the cosmological model of the Universe has been presented in $f(Q)$ gravity and the parameters are constrained from the cosmological data sets. At the beginning, we have employed a well motivated form of $f(Q) = \alpha + \beta Q^{n}$, where $\alpha$, $\beta$ and $n$ are model parameters. We have obtained the Hubble parameter in redshift with some algebraic manipulation from the considered form of $f(Q)$. Then we parameterize with the recent $Hubble$ data and $Pantheon + SHOES$ data using $\textit{MCMC}$ analysis. We validate our obtained model parameter values with $\textit{BAO}$ data set. A parametrization of the cosmographic parameters shows the early deceleration and late time acceleration with the transition at $z_t\approx0.75$. The $Om(z)$ diagnostics gives positive slope which shows that the model in the phantom phase. Also the current age of Universe has been obtained as, $t_{0} = 13.85~~Gyrs$. Based on the present analysis, it indicates that the $f(Q)$ gravity may provide an alternative to dark energy for addressing the current cosmic acceleration.
\end{abstract}

\maketitle
\textbf{Keywords}: $f(Q)$ gravity, Accelerating Universe, $Hubble$ data, $Pantheon + SHOES$ data, Phantom phase.

\section{Introduction}
The most well known theory for gravitational interactions is General Relativity (GR). Based on Riemannian geometry, the Levi-Civita connection has been used to describe GR. The Ricci curvature $R$ has been taken into account as the space time unit and the geometry is free from torsion and nonmetricity. The research devoted to its modification and extension is basically for two reasons: (i) the modified gravity hypothesis arises from cosmological grounds and it is an efficient way to explain the expansion of the Universe over time \cite{Capozziello2011, Saridakis2021} and; (ii) for a purely theoretical motivation. It aims at enhancing the renormalizability of GR with the further goal of achieving a quantum gravity theory\cite{Stelle1977}. Within GR, the observed late time accelerated expansion of the Universe has been modeled by the cosmological constant $\Lambda$ \cite{Riess1998, Perlmutter1999, Spergel2003, Eisenstein2005, Betoule2014, Ade2016, Aghanim2016}. Recently, a cosmological model known as $\Lambda$CDM (Cold Dark Matter) model has been faced with several challenges. There are alternatives to the $\Lambda$CDM, such as modified gravity theories, which modify the long-range gravitational interaction. The gravitational force can also be expressed in terms of other geometries, such as teleparallel gravity \cite{Aldrovandi2014}, in which the force is governed by the torsion $T$ rather than the curvature $R$. Another approach would be the nonmetricity approach, where the nonmetricity $Q$ mediates the gravitational interaction. The nonmetricity $Q$ is curvature-free and torsion-free and the gravity based on the nonmetrtcity is known as the symmetric teleparallel gravity \cite{Nester1998}.

The $f(Q)$ theory of gravity is another modified theory of gravity which recently attracted a lot of attention \cite{Jimenez2018}. To test whether the cosmic behaviour deviates from GR, standard cosmological approaches assume a specific form of relative action. Using the redshift scale factor relation, Lazkoz et al. reformulated the $f(Q)$ gravity and analyse their models with the cosmological data sets \cite{Lazkoz2019}. It has been explained that how dark energy can be identified by space time and ultimately the expansion of the Universe. Based on the current observational bounds of the cosmographic parameters, another reconstruction of $f(Q)$ gravity has been performed using the numerical inversion procedure \cite{Capozziello2022}. Jimenez et al. \cite{Jimenez2020} analyses the behaviour of the cosmological perturbations and discuss the potential strong coupling problem of the maximally symmetric background caused by the discontinuity in the number of propagating modes. Ayuso et al. \cite{Ayuso2021} focuses on a class of $f(Q)$ theories that has been characterized with the presence of a general power law term. Esposito et al. \cite{Esposito2022} presented a reconstruction algorithm for cosmological models based on $f(Q)$ gravity and  obtained the  exact solutions for Bianchi type I and FLRW space times. Anagnostopoulos et al. \cite{Anagnostopoulos2022} showed that $f(Q)$ gravity can safely pass the Big Bang Nucleosynthesis constraints, and in some cases this can be obtained trivially. Another interesting feature on $f(Q)$ gravity is the study of the effect of connections in the dynamics of FLRW cosmology [Dimakis et al.  \cite{Dimakis2022}]. Hu et al. \cite{Hu2022} have demonstrated the Hamiltonian analysis of $f(Q)$ gravity by fixing the coincident gauge condition and have shown that the $f(Q)$ gravity has eight physical degrees of freedom. The linear cosmological perturbations using $1+3$ covariant gauge-invariant formalism has been analysed by Sahlu et al. \cite{Sahlu2022}. Using a symmetric teleparallel scalar tensor theory containing a non-minimal coupling between the nonmetricity scalar and the scalar field, Bahamonde et al. \cite{Bahamonde2022} studied static spherical symmetric configurations. Further, Lin et al. \cite{Lin2021} have studied the application of $f(Q)$ gravity to spherical symmetric configurations and have shown that its effects can be demonstrated by the external and internal solutions of compact stars.

The asymptotic value of growth index and the varying form of growth index in $f(Q)$ gravity has been analysed in [Khyllep et al. \cite{Khyllep2021}].  Anagnostopoulos et al. \cite{Anagnostopoulos2021} have shown the first evidence for $f(Q)$ gravity that challenges the $\Lambda CDM$ behaviour. Frusciante \cite{Frusciante2021} has investigated the impact on cosmological observable of $f(Q)$ gravity. The behaviour of scale factor controlled dynamical parameters have been extensively analysed for the $f(Q)$ gravity in [Narawade et al. \cite{Narawade2022}]. Albuquerque et al. \cite{Albuquerque2022} have analyzed the  phenomenological $f(Q)$ models that provides the kind of deviations expected to differentiate it from the $\Lambda CDM$ scenario. The theory of the accelerating expansion is an intrinsic property of the Universe geometry without need of either exotic dark energy or extra fields and the dynamical system method have been investigated in \cite{Lu2019}. In the context of non-minimal matter couplings, Harko et al. \cite{Harko2018} have tested the consistency of the  new geometrical formulation in terms of the $Q$ while implementing in the matter sector.

A further extension of the nonmetricity gravity is the $f(Q,T)$ gravity \cite{Xu2019} that has been essential to the development of model of cosmic acceleration in the late Universe \cite{Pati2021,Agrawal2021,Pati2022}.

We have discussed here some of the recent results obtained from the cosmological models with observational data. New consistency tests for the $\Lambda CDM$ model has been derived and analysed by Seikel et al. \cite{Seikel2012} that has been formulated in terms of $H(z)$. Slowing down of acceleration of the Universe has been shown by parametrizing the equation of state (EoS) parameter of dark energy and using four sets of Type Ia supernovae data \cite{Magana2014}. Wei et al. \cite{Wei2007} have confronted ten cosmological models with $Hubble$ data set to validate their models. Farooq et al. \cite{Farooq2017} have restricted the spatially flat and curved dark energy models using the Hubble parameter $H(z)$ at the redshift range $(0.07,2.36)$. It is worthy to mention here that recently constrained value of EoS parameter from different cosmological observations are: $\omega = -1.29_{-0.12}^{+0.15}$\cite{Valentino2016}, $\omega = -1.3$\cite{Vagnozzi2020} and $\omega = -1.33_{-0.42}^{+0.31}$\cite{Valentino2021}. A parametric reconstruction of the jerk parameter $j$ which is the dimensionless representation of the third order time derivative of the scale factor has been discuss by Mukherjee et al. \cite{Mukherjee2016}. For the bulk viscous anisotropic Universe, the present value of the Hubble parameter obtained by using the $Hubble$ data, $Pantheon$ and joint $Hubble$ and $Pantheon$ data are respectively $H_{0}=69.39\pm 1.54kms^{-1}Mpc^{-1}$, $70.016\pm 1.65kms^{-1}Mpc^{-1}$ and $69.36\pm 1.42kms^{-1}Mpc^{-1}$.

The paper has been organised as: In Sec.\ref{Sec:II} we have derived the cosmological equations for the spatially flat FLRW space time in $f(Q)$ gravity. Based on Markov chain Monte Carlo (MCMC) procedure, we have done the statistical analysis of the $f(Q)$ model in Sec.\ref{Sec:III}. The baryon acoustic oscillation ($\textit{BAO}$) data set used to obtain the parameters in Sec.\ref{Sec:IV}. The cosmographic parameters have been presented in Sec.\ref{Sec:V} and $Om(z)$ diagnostic in Sec.\ref{Sec:VI}. The results obtained and the conclusions are given in Sec.\ref{Sec:VII}.

\section{Field Equation for $f(Q)$ Gravity}\label{Sec:II}
The action of $f(Q)$ gravity can be given as \cite{Jimenez2018},
\begin{equation}\label{eq:1}
S = \int \left[ \frac{1}{2}f(Q)+\mathcal{L}_{m} \right]\sqrt{-g}dx^{4},
\end{equation}
where $f(Q)$ is an arbitrary function of nonmetricity $Q$, $\mathcal{L}_{m}$ be the matter Lagrangian density and $g$ is the determinant of the metric tensor $g_{\mu\nu}$ . The nonmetricity tensor is give by,
\begin{equation}\label{eq:2}
Q_{\alpha\mu\nu} = \nabla_{\alpha}g_{\mu\nu}.
\end{equation}
Here, both the curvature and torsion are vanishing and the geometry has been determined by the nonmetricity. The nonmetricity tensor is characterized by two independent traces, namely $Q_{\mu} = Q_{\mu~~\alpha}^{~~\alpha}$ and $\tilde{Q}^{\mu} = Q_{\alpha}^{~~\mu\alpha}$ depending on the order of contraction. The nonmetricity scalar which is a quadratic combination, invariant under local linear transformations has been given as,
\begin{equation}\label{eq:3}
Q = -\frac{1}{4}Q_{\alpha\beta\mu}Q^{\alpha\beta\mu} + \frac{1}{2}Q_{\alpha\beta\mu}Q^{\beta\mu\alpha} + \frac{1}{4}Q_{\alpha}Q^{\alpha} - \frac{1}{2}Q_{\alpha}\tilde{Q}^{\alpha}.
\end{equation}
\begin{widetext}
The superpotential of the model is,
\begin{equation}\label{eq:4}
P^{\alpha}_{~~\mu \nu} \equiv -\frac{1}{4}Q^{\alpha}_{~~\mu \nu} + \frac{1}{4}\left(Q^{~~\alpha}_{\mu~~\nu} + Q^{~~\alpha}_{\nu~~\mu}\right) + \frac{1}{4}Q^{\alpha}g_{\mu \nu} - \frac{1}{8}\left(2 \tilde{Q}^{\alpha}g_{\mu \nu} + {\delta^{\alpha}_{\mu}Q_{\nu} + \delta^{\alpha}_{\nu}Q_{\mu}} \right).
\end{equation}
The gravitational field equation can be obtained by varying action \eqref{eq:1}
 with respect to the metric tensor as in \cite{Jimenez2018}
 \begin{equation}\label{eq:5}
 \frac{2}{\sqrt{-g}}\nabla_{\alpha}(\sqrt{-g}f'P^{\alpha}_{~\mu \nu}) + \frac{1}{2}g_{\mu \nu}f + f'(P_{\mu \nu \beta}Q^{~~\alpha \beta}_{\nu} - 2Q_{\alpha \beta \mu}P^{\alpha \beta}_{~~~\nu}) = -T_{\mu \nu},
 \end{equation}
 \end{widetext}
 where $f' = \frac{df}{dQ}$ and $T_{\mu \nu} = -\frac{2}{\sqrt{-g}}\frac{\delta \left( \sqrt{-g}\mathcal{L}_{m}\right)}{\delta g^{\mu \nu}}$. We consider isotropic, homogeneous and spatially flat FLRW space time,
 \begin{equation}\label{eq:6}
 ds^{2} = -dt^{2} + a^2(t)(dx^2+dy^2+dz^2),
 \end{equation}
 where $a(t)$ is the scale factor that describes the expansion rate in the spatial directions and the Hubble parameter, $H = \frac{\dot{a}}{a}$, an over dot represents derivative in cosmic time. The energy momentum tensor is that of perfect fluid and can be expressed as,
 \begin{equation} \label{eq:7}
T_{\mu\nu} = (\rho + p)u_{\mu}u_{\nu} + pg_{\mu\nu},  
 \end{equation}
where $\rho$ and $p$ are respectively denotes the energy density and pressure of the fluid. Now the field equations \eqref{eq:5} can be obtained as,
 \begin{eqnarray}
 Qf'-\frac{f}{2} &=& \rho, \label{eq:8} \\
 (2Qf''+f')\dot{H} &=& -\frac{1}{2}(\rho + p). \label{eq:9}
 \end{eqnarray} 
Here we consider, $8\pi G = c = 1$. The continuity equation can be given as ,
\begin{equation}\label{eq:10}
 \dot{\rho} = -3H(\rho + p),  
\end{equation}
which is consistent with the cosmological equations in case of standard matter.
 
\section{The model}\label{Sec:III}
To obtain the cosmological parameters, a well defined form of $f(Q)$ is needed, so that the analysis of the cosmological model can be performed. Capozziello et al. \cite{Capozziello2022} have assumed the Pade's approximation to compute a numerical reconstruction of the cosmological observables up to high redshifts. This can reduce the convergence issues associated with standard cosmographic methods and provides an effective method for describing cosmological observables up to high redshifts. Using the relation $Q = 6H^{2}$ to reconstruct $f(Q)$ through a numerical inversion procedure, the best analytical match to numerical outcomes is given by the function, 
\begin{equation}\label{eq:11}
f(Q) = \alpha + \beta Q^{n}, 
\end{equation}
where $\alpha$, $\beta$ and $n$ are free model parameters, the parameter $n > 1$ is a real number responsible for the accelerating phase in the early universe \cite{Capozziello2022a}. There are several $f(Q)$ models which shares the same background evolution as in $\Lambda CDM$, while leaving precise and measurable effects on cosmological observable. But with the increase in redshift, the function $f(Q) = \alpha + \beta Q^{n}$ suggests small deviations from the $\Lambda CDM$ model. For $\alpha = 0$, $\beta = 1$ and $n = 1$, one can recover the aforementioned class of theories with the same background evolution as in GR.

\subsection{Observational Analysis}
The three model parameters present in eqn. \eqref{eq:11} will regulate the dynamical behaviour of the model. The value of the model parameters $\alpha$, $\beta$ and $n$ to be chosen in  such a way that the deceleration parameter attain the value, $q_{0} = -0.54$ \cite{Hernandez2020, Garza2019,Akarsu2019}. From the relationship between scale factor and redshift, $a(t)=\frac{1}{1+z}$, one can get, $H=\frac{-\dot{z}}{1+z}$. Now, from eqns. \eqref{eq:8} and \eqref{eq:9}, we obtain
\begin{equation}\label{eq:12}
\dot{H}= \frac{-(1+\omega)}{4}\frac{2Qf'-f}{2Qf''+f'}.
\end{equation}
Subsequently,
\begin{equation}\label{eq:13}
\frac{dH}{dz} = \frac{(1+\omega)}{4H(1+z)}\frac{2Qf'-f}{2Qf''+f'},
\end{equation}
Substituting eqn. \eqref{eq:11} in eqn. \eqref{eq:13}, we get
\begin{equation}\label{eq:14}
H(z) = H_{0}\left[ \sqrt{\frac{\alpha + (1+z)^{3(1+\omega)}}{2\beta n 6^{n}-\beta 6^{n}}} \right]^{\frac{1}{n}},
\end{equation}
where $H_{0}$ is the present value of the Hubble parameter. Now, our aim is to put constraint on the parameters $\alpha$, $\beta$, $n$ and $\omega$ using the cosmological data sets. For matter dominated phase ($\omega=0$) and dark energy dominated phase ($\omega$ is constant), we obtain the following relation for the Hubble parameter, which is analogous to epsilon model given in \cite{Lemos2018}, 
\begin{widetext}
\begin{equation}\label{eq:15}
H(z) = H_{0}\left[ \sqrt{\left(1-\frac{\alpha + 1}{\beta (2n-1) 6^{n}}\right)(1+z)^{3} +  \frac{1}{\beta (2n-1) 6^{n}}(1+z)^{3(1+\omega)} + \frac{\alpha}{\beta (2n-1) 6^{n}}}  \right]^{\frac{1}{n}}.
\end{equation}
\end{widetext}

\subsubsection{\textit{$Hubble$ Dataset}}
Through estimations of their differential evolution, early type galaxies provide Hubble parameter measurements. The process of compilation of such observations is known as the cosmic chronometers. The list of $32$ data points of Hubble parameter in the redshift range $0.07 \leq z \leq 1.965$ with errors are given in TABLE-\ref{table:IV}. By minimizing the chi-square value, we determine the mean values of the model parameters $\alpha$, $\beta$, and $n$. The Chi-square function can be given as, 
\begin{equation}\label{eq:16}
\chi_{OHD}^{2}(p_{s}) = \sum_{i=0}^{32}\frac{\left[H_{th}(z_{i}, p_{s}) - H_{obs}(z_{i})\right]^{2}}{\sigma_{H}^{2}(z_{i})},
\end{equation}
where $H_{obs}(z_{i})$ represents the observed Hubble parameter values, $H_{th}(z_{i}, p_{s})$ represents the Hubble parameter with the model parameters and $\sigma_{H}^{2}(z_{i})$ is the standard deviation.

\subsubsection{\textit{$Pantheon + SHOES$ Dataset}}
The $Pantheon + SHOES$ sample data set consists  of 1701 light curves of 1550  distinct Type Ia supernovae (SNe Ia) ranging in redshift from $z = 0.00122$ to $2.2613$ \cite{Brout2022}. The model parameters are to be fitted by comparing the observed and theoretical value of the distance moduli. The distance moduli can be defined as,
\begin{equation} \label{eq:17}
\mu(z, \theta) = 5log_{10}[d_{L}(z,\theta)] + \mu_{0},
\end{equation}
where $\mu_{0}$ is the nuisance parameter and $d_{L}$ is the dimensionless luminosity distance defined as,
\begin{equation}\label{eq:18}
d_{L}(z) = (1+z)\int_{0}^{z}\frac{d\tilde{z}}{E(\tilde{z})} ,
\end{equation}
where $E(z) = \frac{H(z)}{H_{0}}$ is the dimensionless parameter and $\tilde{z}$ is variable change to define integration from $0$ to $z$. The $\chi^{2}$ is given by,
\begin{equation}\label{eq:19}
\chi^{2}_{SN}(z,\theta) = \sum_{i=1}^{1701}\frac{\left[\mu(z_{i},\theta)_{th} - \mu(z_{i})_{obs}\right]^{2}}{\sigma^{2}_{\mu}(z_{i})},
 \end{equation}
 where $\sigma^{2}_{\mu}(z_{i})$ is the standard error in the observed value. In order to calculate $\chi^{2}_{SN}$, we use the fact that the SNIa data set corresponds to redshifts below 2 so that we can neglect the contribution from radiation in Einstein's equations.
 
 \subsubsection{\textit{MCMC} Analysis}
In order to obtain the tight constraints on the parameters of the $f(Q)$ model, the  \textit{MCMC} analysis will be used to implement the test. This analysis will produce proficient fits of $\alpha$, $\beta$, $\omega$ and $n$ upon minimization of a total $\chi^{2}$. Also, this analysis will produce selection criteria, which will allow us to draw some conclusions. The panels on the diagonal in corner MCMC plot shows the $1-D$ curve for each model parameter obtained by marginalizing over the other parameters, with a thick line curve to indicate the best fit value. The off diagonal panels show $2-D$ projections of the posterior probability distributions for each pair of parameters, with contours to indicate $1-\sigma$ (Blue) and $2-\sigma$ (Light Blue) regions. The best fit values of  $\alpha$, $\beta$, $\omega$ and $n$ are obtained from the triangle plot FIG-\ref{fig:I}(left panel) through $Hubble$ data and in triangle plot Fig-\ref{fig:I}(right panel) using $Pantheon + SHOES$ data with $1-\sigma$ and $2-\sigma$ confidence intervals. All the obtained values are listed in Table \ref{table:I}. In  Fig-\ref{fig:II}, the curve for distance modulo has been given as the distance modulo can also expressed in Hubble parameter eqn.\eqref{eq:15}. Here, the solid red line passed in the middle through the error bar plots [Fig-\ref{fig:II}], where we used best fit values from Table \ref{table:I} to plot the error bar plot. Also, In Fig-\ref{fig:III}, we have shown the error bar plots of $H(z)$ [upper panel] and $H(z)/(1 + z)$ [Lower panel] using the best fit values obtained in Table \ref{table:I}.
  \begin{widetext} The dotted line represents the $\Lambda$CDM line and the solid red line represents the best fit curve for Hubble rate. It can be observed that in both the figures, the solid red line is traversing at the middle of the error bars. We have marginalized value of $H_{0}$ as $69.5^{+2.3}_{-1.9} kms^{-1}Mpc^{-1}$ and $70.7\pm2.7 kms^{-1}Mpc^{-1}$ respectively with Hubble data and $Pantheon + SHOES$ data sets. The details values of the parameters are given in Table-\ref{table:I}. For the subsequent study, we follow $H_0=70.7\pm2.7 kms^{-1}Mpc^{-1}$.

\begin{figure} [!htb]
\centering
\includegraphics[width=8.92cm,height=10cm]{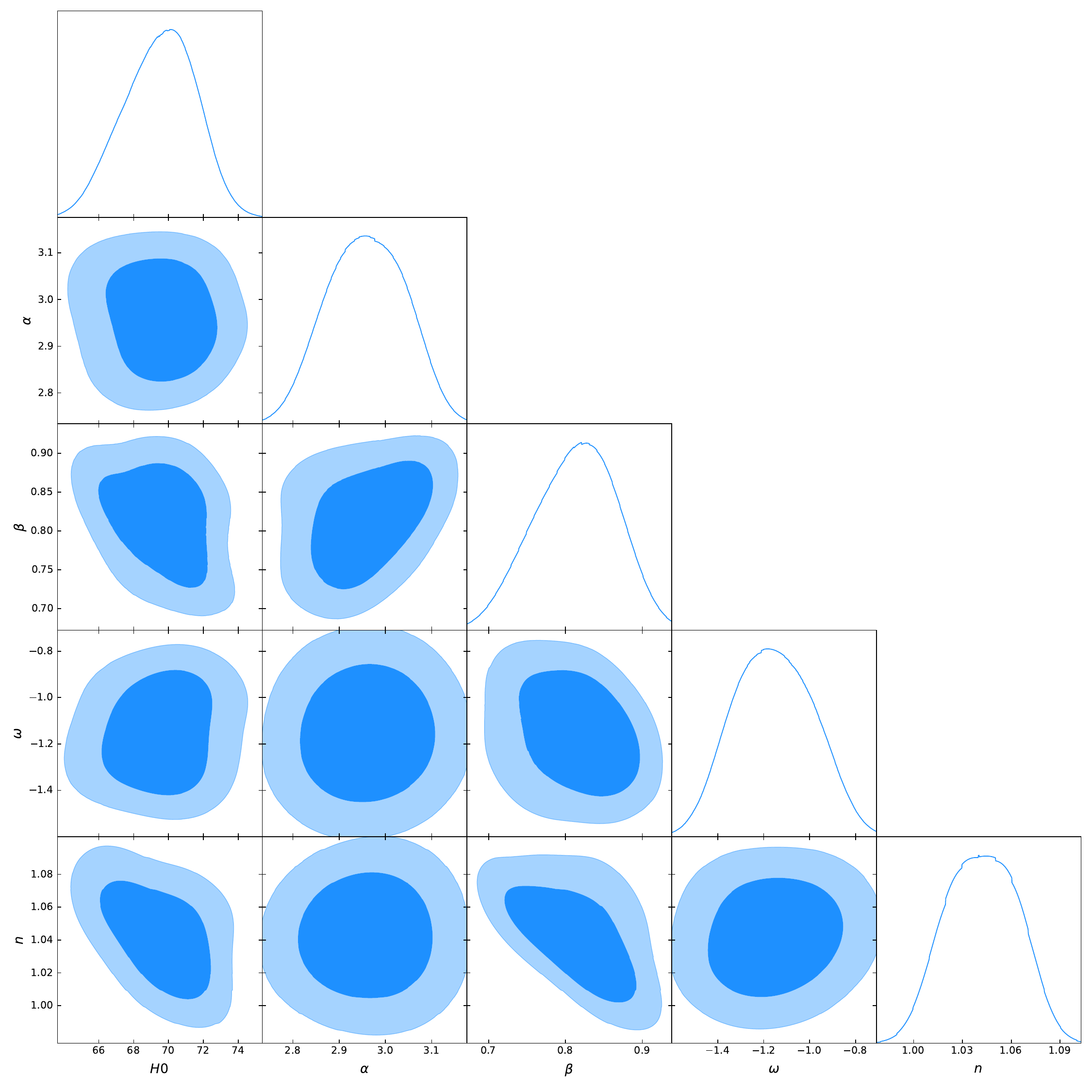}
\includegraphics[width=8.92cm,height=10cm]{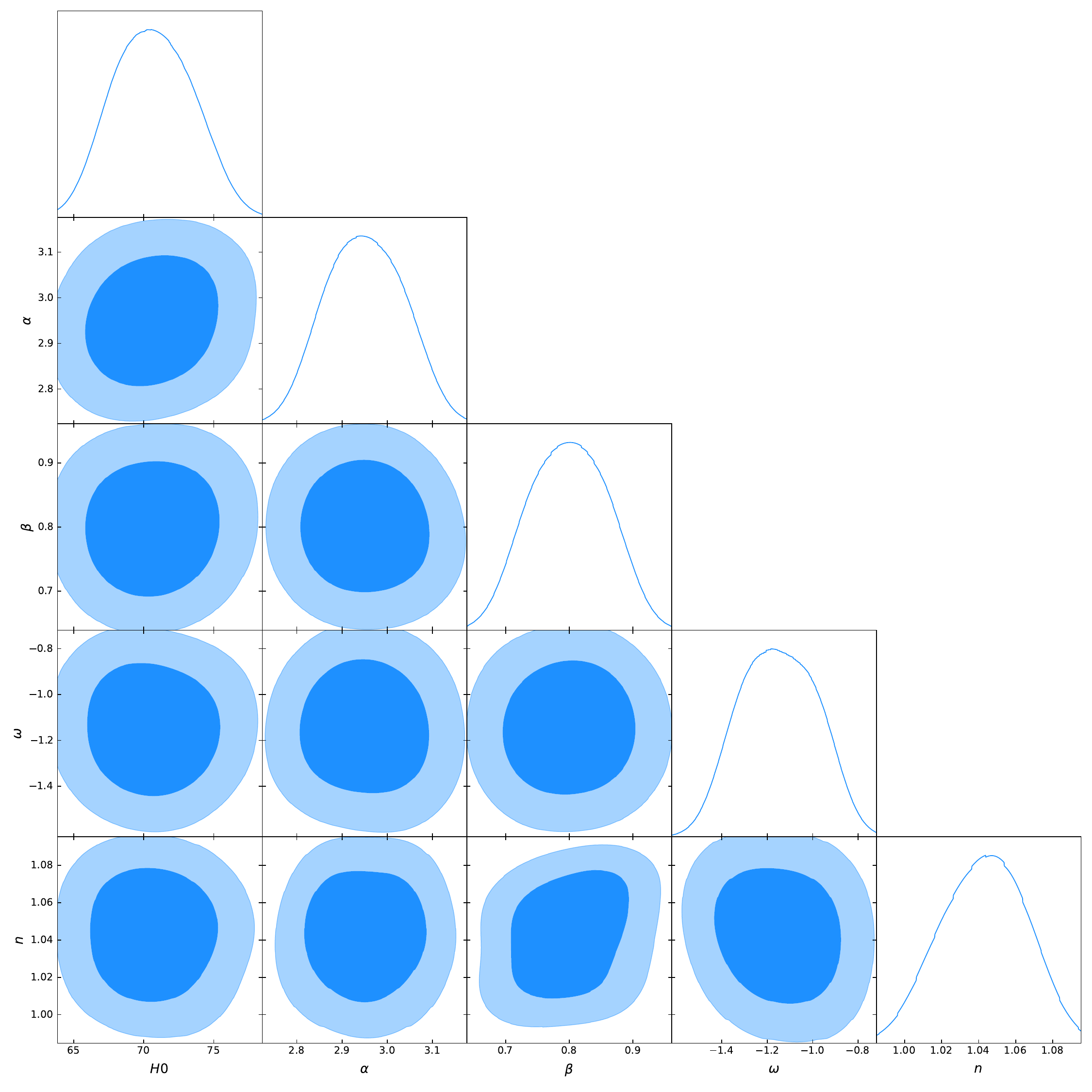}
\caption{The marginalized constraints on the coefficients of $H(z)$ from 32 data points of $Hubble$ data[\textbf{left panel}] and $Pantheon + SHOES$ sample [\textbf{right panel}] with $1\sigma$ and $2\sigma$ confidence intervals.} \label{fig:I}
\end{figure}

\begin{figure}[!htb]
\centering
\includegraphics[scale=0.5]{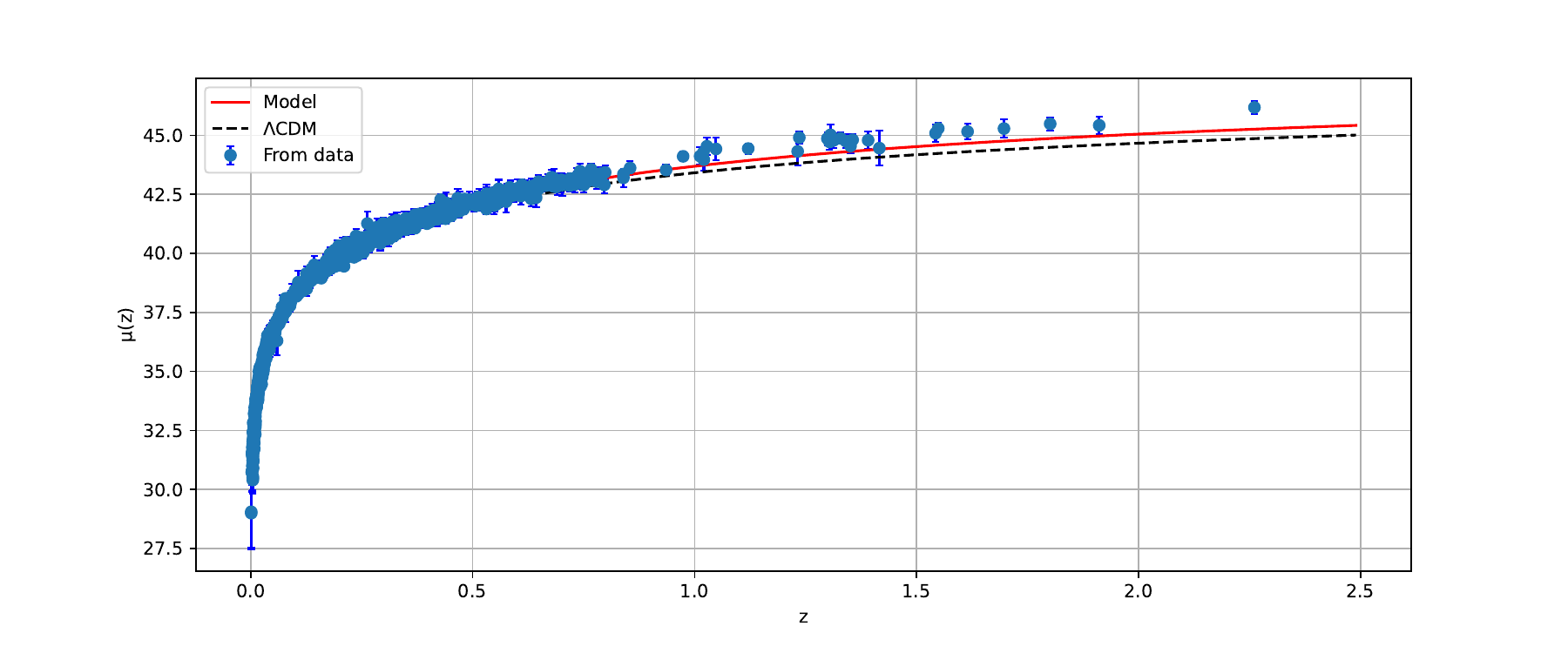}
\caption{$\mu(z)$ in redshift in $Pantheon + SHOES$ data set. The points with error bars indicate the observed $Hubble$ data. The solid red line represents for the model.}
\label{fig:II}
\end{figure}
\begin{figure}[H]
\centering
\includegraphics[scale=0.5]{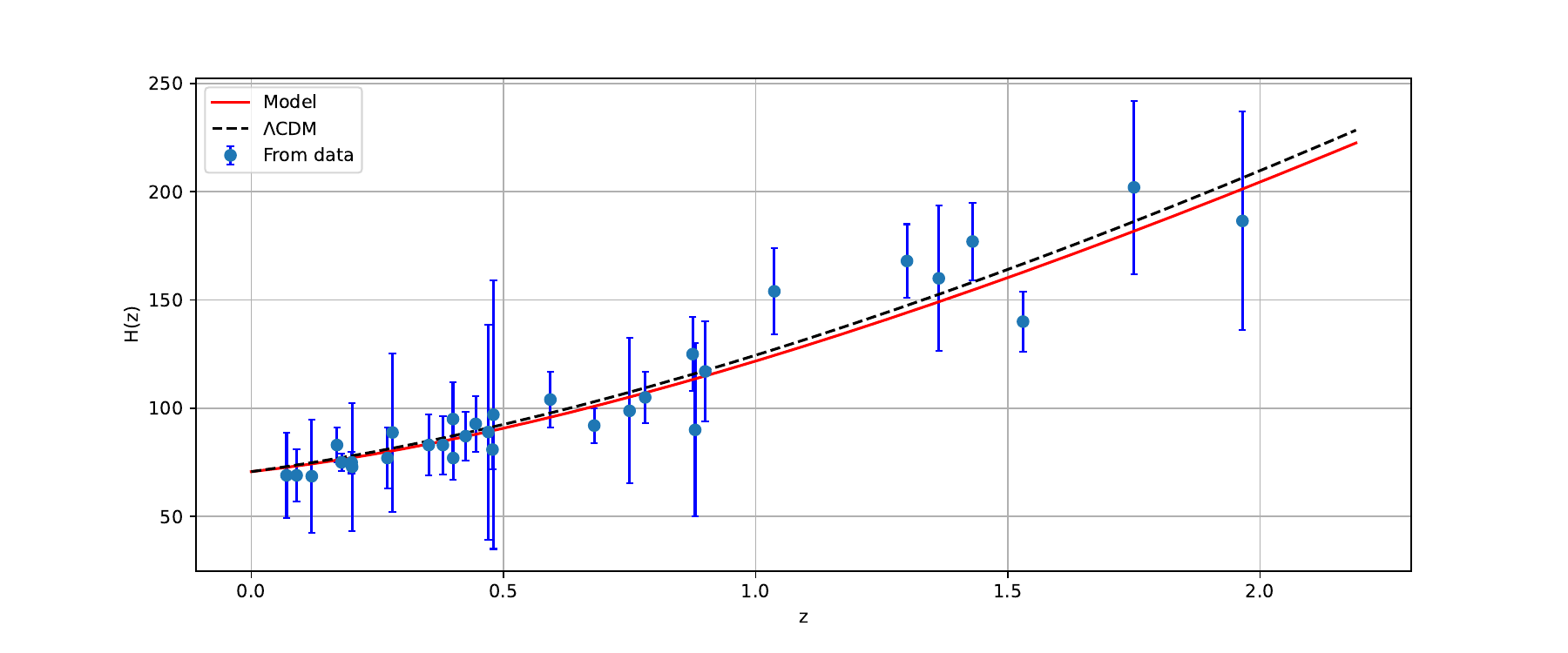}
\includegraphics[scale=0.5]{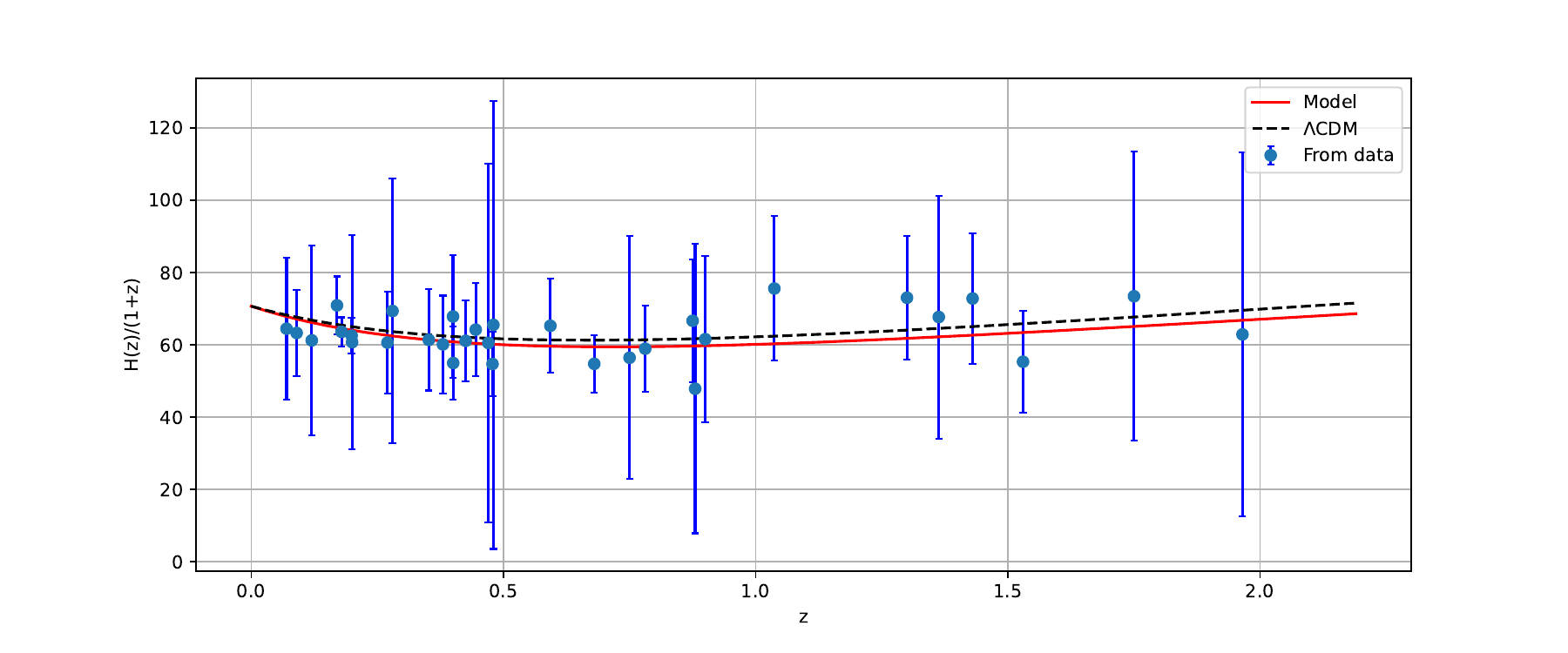}
\caption{$H(z)$ [{\bf upper panel}] and $H(z)/(1 + z)$ [{\bf lower panel}] in redshift from 32 data points of $Hubble$ data [Ref. TABLE \eqref{table:IV}]. For the $Hubble$ data: points with blue color error bars and for the model: solid red line. }
\label{fig:III}
\end{figure}
\end{widetext}

\begin{table}[H]
\renewcommand\arraystretch{2}
\centering
\caption{The marginalized constraining results of the parameters using $Hubble$ and $Pantheon + SHOES$ data}
    \label{table:I}
\begin{tabular}{|c|c|c|}
\hline\hline
~Parameters~ &~~$Hubble$ dataset~~&~~$Pantheon + SHOES$ dataset~~  \\
\hline
$H_{0}$ & $69.5_{-1.9}^{+2.3}$ &  $70.7\pm 2.7$\\
\hline
$\omega$ & $-1.16\pm0.17$ &  $-1.15\pm 0.17$\\
\hline
$\alpha$ & $2.96\pm 0.082$ &  $2.951\pm 0.082$\\
\hline
$\beta$ & $0.813\pm 0.059$ &  $0.796\pm 0.0059$ \\
\hline
$n$ & $1.041 \pm 0.021$ &  $1.043\pm 0.022$\\
\hline\hline
\end{tabular}
\end{table}

\section{Baryon acoustic oscillations and cosmic microwave background}\label{Sec:IV}
In $\Lambda$CDM observations, the $\textit{BAO}$ standard ruler measurements are self consistent with $\textit{CMB}$ observations, as demonstrated by several galaxy surveys. As mentioned, our model has a similar background to $\Lambda$CDM but shows slight deviations at high redshift. So we can obtain stringent constraints on cosmological parameters by using this measurements. To map distance-redshift relationships, measuring the $\textit{BAO}$s in large scale clustering patterns of galaxies is a promising technique. Also, $\textit{BAO}$ provides an independent way to measure the expansion rate of the Universe and also can describe the rate of change of expansion throughout the evolution history.\\

\subsubsection{\textit{BAO}}
The angular diameter distance through the clustering perpendicular to the line of sight can be measured using the $\textit{BAO}$ signals. Moreover, the expansion rate of the Universe $H(z)$ can be measured by the clustering along the line of sight. At the photon decoupling epoch, the comoving sound horizon can be defined as,
\begin{equation}\label{eq:20}
r_{s}(z_{*}) = \frac{c}{\sqrt{3}}\int_{0}^{1/(1+z_{*})}\frac{d\tilde{z}}{\tilde{z}^{2}H(\tilde{z})\sqrt{1+\tilde{z}\left(3\Omega_{b0}/4\Omega_{\gamma0}\right)}},
\end{equation}
where, $\Omega_{b0}$ and $\Omega_{\gamma0}$ respectively represent the present value of the baryon and photon density parameter and $z_{*}$ is the redshift of photon decoupling. According to WMAP7 \cite{Jarosik2011}, here we use, $z_{*} = 1091$. The expression for the comoving angular-diameter distance [$d_{A}(z_{*})$] and the dilation scale [$D_{V}(z)$] are respectively,
\begin{eqnarray}
d_{A}(z_{*}) = \int_{0}^{z_{*}}\frac{d\tilde{z}}{H(\tilde{z})}, \nonumber\\
D_{V}(z) = \left[\frac{(d_{A}(z))^{2}cz}{H(z)}\right]^{\frac{1}{3}}. \label{eq:21}
\end{eqnarray}
The epoch at which baryons were released from photons called as drag epoch $(z_{d})$. At this epoch, the photon pressure is no longer able to avoid gravitational instability of the baryons. We use the value, $z_{d} = 1020$ \cite{Komatsu2009}.

\subsubsection{\textit{CMB}}
The $\textit{CMB}$ is leftover radiation from the Big Bang or the time when the Universe starts evolution. In order to confront the dark energy models to $\textit{CMB}$ data, the distance priors method is more appropriate \cite{Wang2007, Wright2007}. Using the $\textit{CMB}$ temperature power spectrum, this method measures two distance ratios:
\begin{itemize}
\item[(i)] The acoustic scale, measures the ratio of the angular diameter distance to the decoupling epoch. At decoupling epoch, it also also measures the size of the comoving sound horizon. This first distance ratio can be expressed as,
\begin{equation*}
l_{A} = \pi\frac{d_{A}(z_{*})}{r_{s}(z_{*})}.
\end{equation*}
\item[(ii)] The second one is at decoupling time, the ratio of angular diameter distance and the Hubble ratio, called the shift parameter. This can be expressed as,
\begin{equation*}
R = \sqrt{\Omega_{m}H_{0}^{2}}r(z_{*}).
\end{equation*}
\end{itemize}
The acoustic scale is used to obtain the $\textit{BAO/CMB}$ constraints. Combining these results with the WMAP7-year \cite{Jarosik2011} and WMAP9-year \cite{Bennett2013} the value $l_{A} = 302.44\pm0.80$ and $l_{A} = 302.35\pm 0.65$ respectively. Percival et al. \cite{Percival2010} measured $\frac{r_{s}(z_{d})}{D_{V}(z)}$, at $z=0.2$ and $z =0.35$. The 6dF Galaxy Survey also reported a new measurement of $\frac{r_{s}(z_{d})}{D_{V}(z)}$ at $z =0.106$  \cite{Beutler2011} and WiggleZ team \cite{Blake2011} obtained results at $z=0.44$, $z=0.60$ and $z=0.73$ (see Table \ref{table:II}). By using the WMAP 7 \cite{Jarosik2011}, recommended values for $r_{s}(z_{d})$ and $r_{s}(z_{*})$ we get, $\frac{r_{s}(z_{d})}{r_{s}(z_{*})} = 1.045\pm0.016$. The BAO/CMB constraints $\frac{d_{A}(z_{*})}{D_{V}(z_{BAO})}$ also exhibited in Table \ref{table:II}, along with the values for $\frac{d_{A}(z_{*})}{D_{V}(z_{BAO})}\frac{r_{s}(z_{d})}{r_{s}(z_{*})}$. Now we can write the $\chi^{2}$ for the $\textit{BAO/CMB}$ analysis as, 
\begin{equation*}
\chi_{\textit{BAO/CMB}}^{2} = X^{T}C^{-1}X,
\end{equation*}
where $X$ depends on the survey considered is given by, 
\( X= 
\renewcommand\arraystretch{1.4}
\begin{pmatrix}
\frac{d_{A}(z_{*})}{D_{V}(0.106)}-30.95 \\
\frac{d_{A}(z_{*})}{D_{V}(0.200)}-17.55 \\
\frac{d_{A}(z_{*})}{D_{V}(0.350)}-10.11 \\
\frac{d_{A}(z_{*})}{D_{V}(0.440)}-8.44 \\
\frac{d_{A}(z_{*})}{D_{V}(0.600)}-6.69 \\
\frac{d_{A}(z_{*})}{D_{V}(0.730)}-5.45
\end{pmatrix}
\)\\ and the inverse of covariance matrix $C$ is given by \cite{Giostri2012}, \begin{widetext}  
\begin{equation*}
 C^{-1} = 
\renewcommand\arraystretch{1.3}
\begin{pmatrix}
0.48435 & -0.101383 & -0.164945 & -0.0305703 & -0.097874 & -0.106738 \\
-0.101383 & 3.2882 & -2.45497 & -0.0787898 & -0.252254 & -0.2751 \\
-0.164945 & -2.45497 & 9.55916 & -0.128187  & -0.410404 & -0.447574  \\
-0.0305703 & -0.0787898 & -0.128187 & 2.78728 & -2.75632 & 1.16437 \\
-0.097874 & -0.252254 & -0.410404 & -2.75632 & 14.9245 & -7.32441 \\
0.106738 & -0.2751 & -0.447574 & 1.16437 & -7.32441 & 14.5022 \\
\end{pmatrix}
\end{equation*}

\begin{table}[H]
\renewcommand\arraystretch{2}
\centering
\caption{$\textit{BAO}$ measurements at six different redshifts are now included in the most recent $\textit{BAO}$ distance data set.}
    \label{table:II}
\begin{tabular}{|c|c|c|c|c|c|c|}
\hline
$z_{BAO}$ & 0.106 & 0.200 & 0.350 & 0.440 & 0.600 & 0.730  \\
\hline
$\frac{r_{s}(z_{d})}{D_{V}(z_{BAO})}$ & ~$0.336\pm 0.015$~ & ~$0.1905\pm0.0061$~ & ~$0.1097\pm0.0036$~ & ~$0.0916\pm0.0071$~ & ~$0.0726\pm0.0034$~ & ~$0.0592\pm0.0032$~ \\
\hline
$\frac{d_{A}(z_{*})}{D_{V}(z_{BAO})}$ & $30.95\pm1.46$ & $17.55\pm0.60$ & $10.11\pm0.37$ & $8.44\pm0.67$ & $6.69\pm0.33$ & $5.45\pm0.31$\\
\hline
~~$\frac{d_{A}(z_{*})}{D_{V}(z_{BAO})}\frac{r_{s}(z_{d})}{r_{s}(z_{*})}$~~ & $32.35\pm1.45$ & $18.34\pm0.59$ & $10.56\pm0.35$ & $8.82\pm0.68$ & $6.99\pm0.33$ & $5.70\pm0.31$\\
\hline
\end{tabular}
\end{table}

\begin{figure}[H]
\centering
\includegraphics[scale=0.50]{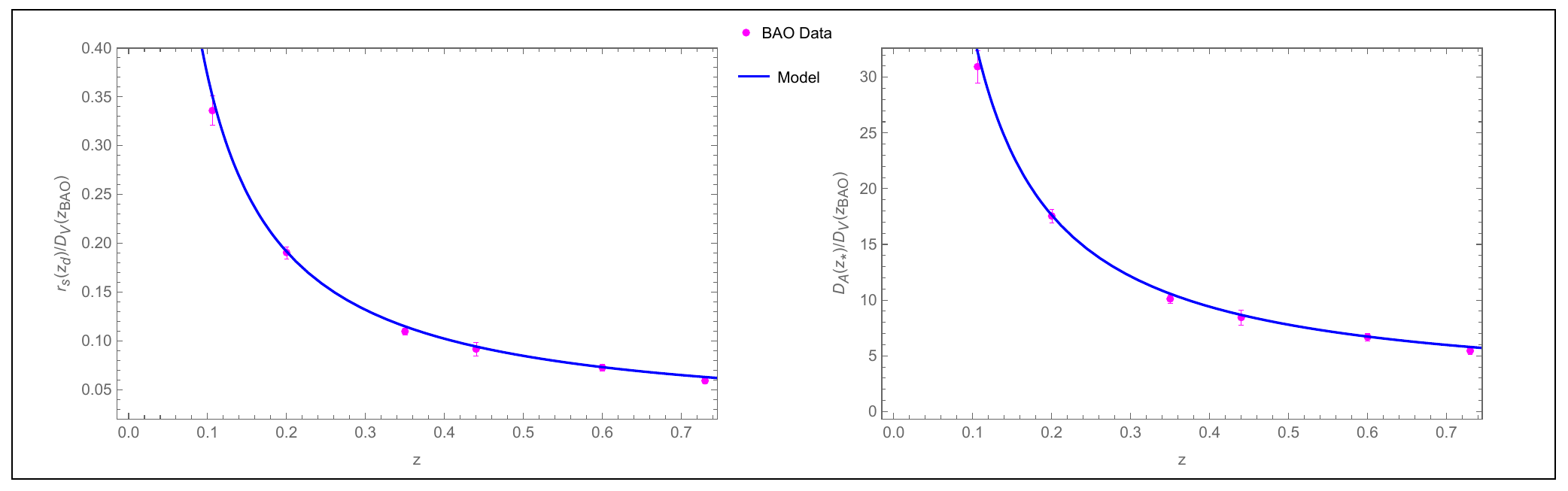}
\caption{Plots of $\frac{r_{s}(z_{d})}{D_{V}(z_{BAO})}$ parameter vs. redshift [{\bf left panel}] and $\textit{BAO/CMB}$ constraints vs. redshift [{\bf right panel}].}
\label{fig:IV}
\end{figure}
\end{widetext}
The plot for the distilled parameter and $\textit{BAO/CMB}$ constraints has been made using the $H_{0}=70.7kms^{-1}Mpc^{-1}$, $\omega = -1.15$, $\alpha = 2.95$, $\beta = 0.796$ and $n = 1.043$ that we obtained in TABLE-\ref{table:I}. As can be seen from the Fig-\ref{fig:IV}, the values we obtained in our results are best suited with the observational values of the $\textit{BAO/CMB}$ data.

\section{Cosmographic Parameters}\label{Sec:V}
 As explained in the cosmographic series\cite{Weinberg1972, Visser2004, Visser2005}, the Hubble parameter $H$ is derived from the scale factor. The second and third derivative of the cosmographic series, the deceleration parameter ($q$) and jerk parameter ($j$) can be determined respectively. The Hubble parameter $H$ is a key element in characterising the evolution of the Universe and it indicates the rate of the expansion of the Universe. For the expanding behaviour of the Universe, the Hubble parameter must be positive. The negative and positive sign of the deceleration parameter respectively gives the information on the accelerating and decelerating behaviour of the Universe. Now, using \eqref{eq:14}, these parameters can be expressed in redshift as,
 \begin{eqnarray}
 q(z) &=& -1+\frac{(1+z)H_{z}(z)}{H(z)}, \nonumber\\
 j(z) &=& q(z)*\left[1+2q(z)\right] + (1+z)*q_{z}(z). \label{eq:22}
 \end{eqnarray}
The interval on the value of $q$ ($q_0$ denotes the present value) describes the behaviour of the Universe as follow:
 \begin{itemize}
 \item[(i)] The Universe experiences expanding behaviour and undergoes deceleration phase for $q_{0}>0$. During this phase, one can obain pressureless barotropic fluid or matter dominated Universe.  However, the results from cosmological observations do not favor positive $q_{0}$. This situation would have occurred during early Universe.
 \item[(ii)] The expanding and accelerating Universe represents for $-1<q_{0}<0$, which is the present status of the Universe. 
 \item[(iii)] The entire cosmological energy budget is dominated by a de Sitter fluid for $q_{0} = -1$. This is the case of inflation during the very early Universe. 
 \end{itemize}
From eqn.\eqref{eq:22}, the present value of the jerk parameter can be, $j_{0} = 2q_{0}^{2}+q_{0}+ q_{z|_0}$. We wish to keep, $-1<q_{0}<-0.5$, which requires $2q_{0}^{2}+q_{0}>0$. Hence, if $q_{0}<-0.5$, then $j_{0}$ is linked to the sign of the variation of $q$. Accordingly the behaviour of this geometrical parameter can be interpreted as follows:
 \begin{itemize}
\item[(i)] when $j_{0}$ is negative, there is no change of the behaviour from the present phase to the accelerated phase. The dark energy influences early time dynamics without any change since the start of evolution.
\item[(ii)] when $j_{0}$ vanishes, the accelerating parameter tends smoothly to a precise value, without any change in its behavior. 
\item[(iii)] when $j_{0}$ is positive, there was a precise point during the evolution when the acceleration of the Universe began. The corresponding redshift can be referred as the transition redshift, at which the effect of dark energy becomes significant. As a consequence,it  indicates the presence of further cosmological resources. In order to constrain the dark energy equation of state, one would need to measure the transition redshift $z_{tr}$ directly. To note here, the slope of the Universe gets changed by changing the sign of $j_{0}$. 
\end{itemize}

\begin{figure}[H]
\centering
\includegraphics[width=90mm]{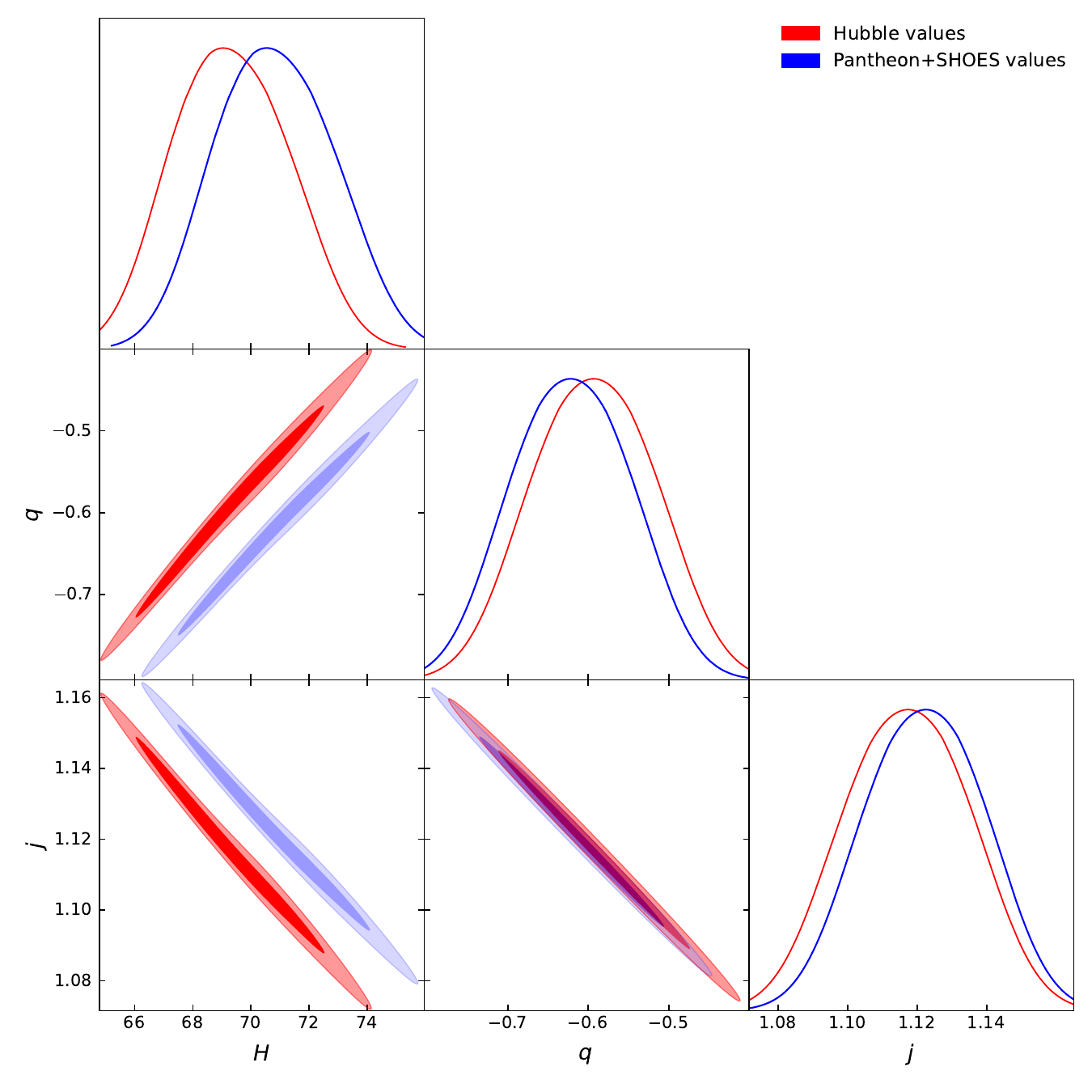} 
\caption{The marginalized constraints on the $H(z)$, $q(z)$ and $j(z)$ using redshift values.} \label{fig:V}
 \end{figure}
 \vspace{9mm}
\begin{widetext}
In Fig.\ref{fig:V}, we have marginalize the $z$-values for the present values of $H(z)$, $q(z)$ and $j(z)$. In Table-\ref{table:III} the best fit present value of $H(z)$, $q(z)$ and $j(z)$ are given, as obtained from the $Hubble$ and $Pantheon + SHOES$ data using marginalized model parameter values.

\begin{figure}[H]
\centering
\includegraphics[width=\textwidth]{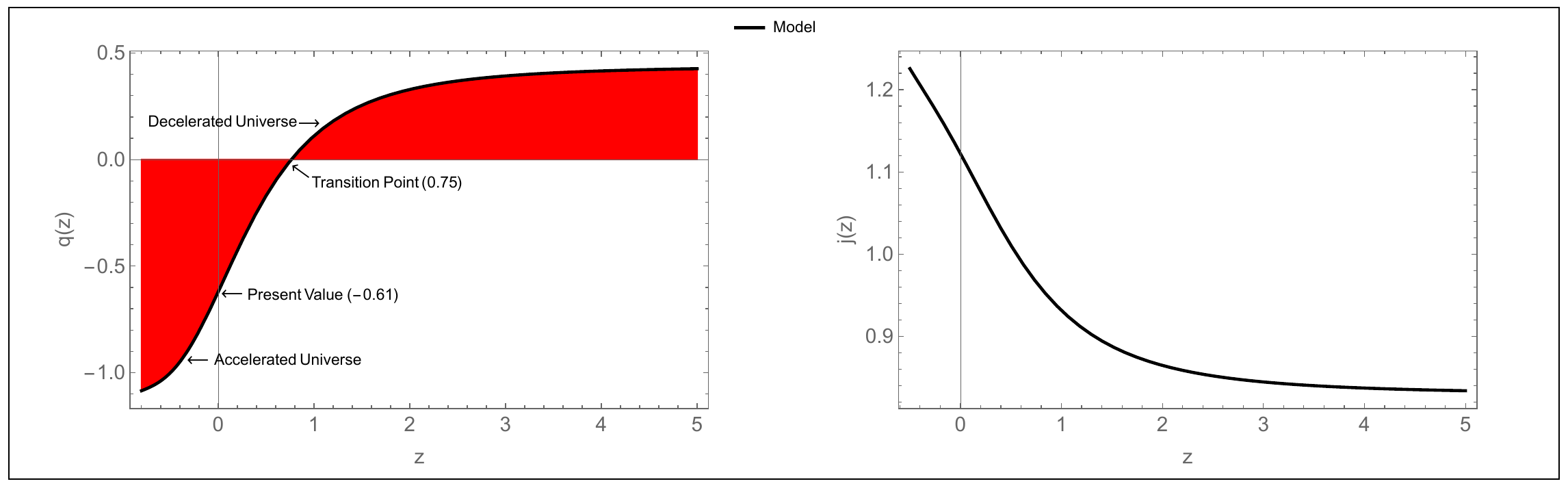}
\caption{Deceleration parameter [{\bf left panel}] and jerk parameter [{\bf right panel}] in redshift. {\bf The parameter scheme:} $\omega = -1.15 $, $\alpha =2.95$, $\beta = 0.796$, $n = 1.043$.}
\label{fig:VI}
\end{figure}

In Fig.\ref{fig:VI} (left panel) the deceleration parameter shows early deceleration to late time acceleration, the transition happens from deceleration to acceleration at $z_{t} \approx 0.75$ and the present value of deceleration parameter obtained as, $q_{0} \approx -0.61$.\end{widetext}\newpage
Recently performed measurements have determined that the value of the deceleration parameter for the current cosmic epoch is within the range of $q_{0} = -0.528_{-0.088}^{+0.092}$ \cite{Gruber2014} and transition from deceleration to acceleration at $z_{t} = 0.60_{-0.12}^{+0.21}$ \cite{Yang2020,Capozziello2015}. From Fig.\ref{fig:VI} (right panel), we can observed that $j>0$, which verify that there exists a transition time when the Universe modifies its expansion.

\begin{table}[H]
\renewcommand\arraystretch{2}
\centering
\caption{The marginalized constraining results of the cosmographic parameters using $Hubble$ and $Pantheon + SHOES$ data.}
    \label{table:III}
\begin{tabular}{|c|c|c|}
\hline\hline
~Parameters~ &~~$Hubble$ dataset~~&~~$Pantheon + SHOES$ dataset~~  \\
\hline
$H(z)$ & $69.5_{-1.9}^{+2.3}$ &  $70.7\pm 2.7$\\
\hline
$q(z)$ & $-0.59\pm 0.07$ &  $-0.61\pm 0.067$\\
\hline
$j(z)$ & $1.117\pm 0.02$ &  $1.122\pm 0.02$ \\
\hline\hline
\end{tabular}
\end{table}

\section{$Om(z)$ Diagnostic and Age of the Universe}\label{Sec:VI}
The $Om(z)$ diagnostic has been introduced as an alternative approach to test the accelerated expansion of the Universe with the phenomenological assumption, EoS, $p=\rho\omega$ filling the universe with the perfect fluid. The $Om(z)$ diagnostic provides a null test to the $\Lambda CDM$ mode \cite{Sahni2008}. Also, there are evidences available in the literature on its sensitiveness with the EoS parameter \cite{Ding2015, Zheng2016, Qi2018}. The nature of $Om(z)$ slope differs between dark energy models because: the positive slope indicates the phantom phase $\omega < -1$, and the negative slope indicates the quintessence region $\omega > -1$. The $Om(z)$ diagnostic can be defined as, 
\begin{equation*}
Om(z) = \frac{E^{2}(z) - 1}{(1+z)^{3}-1},
\end{equation*}
\begin{widetext}
\begin{equation}\label{eq:23}
Om(z) = \frac{\left(\frac{\alpha }{\beta 6^{n}(2n-1)}+(z+1)^3 \left(-\frac{\alpha }{\beta 6^{n}(2n-1)}-\frac{1}{\beta 6^{n}(2n-1)}+1\right)+\frac{(z+1)^{3 (\omega +1)}}{\beta 6^{n}(2n-1)}\right)^{1/n}-1}{(z+1)^3-1}.
\end{equation}

In Fig.\ref{fig:VII} (left panel), $Om(z)$ shows the positive behaviour and that confirms the phantom like behaviour of model.
\begin{figure*}[!htb]
\centering
\minipage{1\textwidth}
\includegraphics[width=\textwidth]{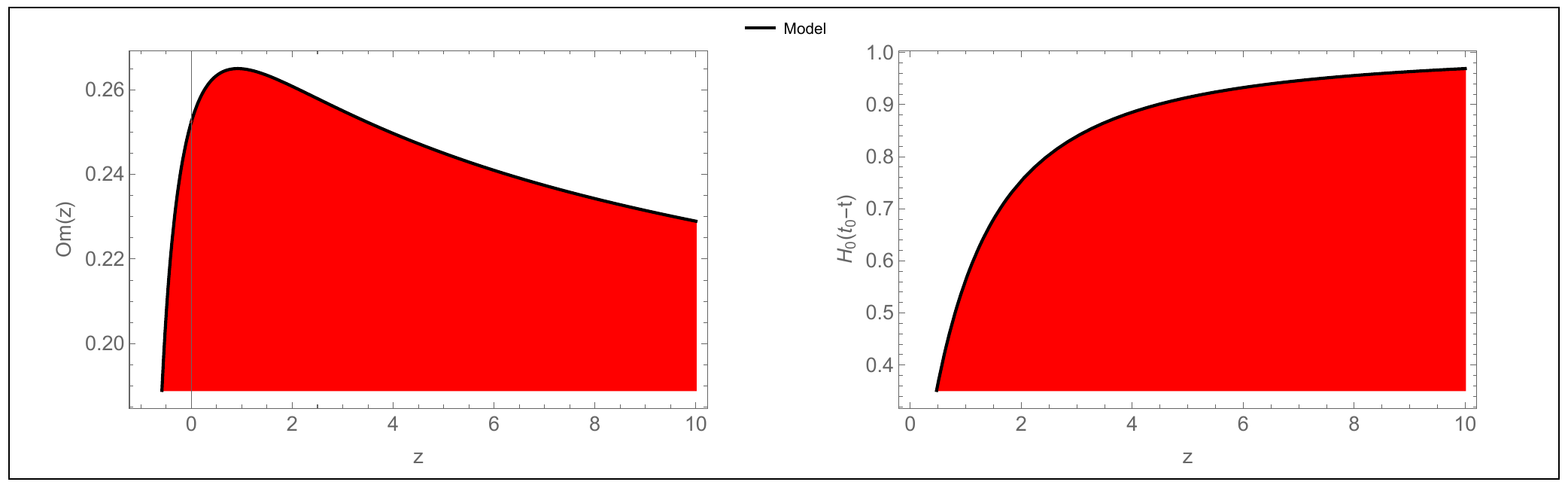}
\endminipage
\caption{Behaviour of $Om(z)$ [{\bf left panel}] and $H_{0}(t_{0}-t)$ [{\bf right panel}] in redshift.{\bf The parameter scheme:} $\omega = -1.15$, $\alpha = 2.95$, $\beta = 0.796$, $n = 1.043$.}
\label{fig:VII}
\end{figure*}
 \end{widetext}
The age-redshift relationship determining the age of the Universe as a function of redshift, $t_{U}(z)$ is given by \cite{Vagnozzi2021}:
\begin{equation}\label{eq:24}
 t_{U}(z) = \int_{z}^{\infty}\frac{d\tilde{z}}{(1+\tilde{z})H(\tilde{z})}.
\end{equation}

 The age of the Universe at any redshift is inversely proportional to $H_{0} = H(z=0)$ as shown in eqn.\eqref{eq:24}. Now, Using eqn.\eqref{eq:15} we shall compute the age of the Universe as,
\begin{eqnarray}
 H_{0}(t-t_{0}) = \int_{0}^{z}\frac{d\tilde{z}}{(1+\tilde{z})E(\tilde{z})}, \nonumber\\
    H_{0}t_{0} = \lim_{z\to\infty}\int_{0}^{z}\frac{d\tilde{z}}{(1+\tilde{z})E(\tilde{z})}.
\end{eqnarray}\label{eq:25}

The behaviour of time with redshift is depicted in Fig.\ref{fig:VII} (right panel). It is found that for infinitely large $z$, $ H_{0}(t_{0}-t)$ converges to $0.9791$. We can use this to calculate the current age of the Universe as $t_{0} = 0.9791H_{0}^{-1} \approx 13.85~~Gyrs$ which is quite close to the age calculated from the Planck result  $t_{0} = 13.78 \pm 0.020~~Gyrs$. So, the results obtained in the model are in consistent with current data.

\section{Conclusion}\label{Sec:VII}
Considering higher power of nonmetricity in the expression of $f(Q)$ and with some algebraic manipulation, we have explicitly obtained the expression for the Hubble parameter in redshift. Then, we have parametrized $H(z)$ with the $Hubble$ and the $Pantheon + SHOES$ data sets. In the redshift range $0.07\leq z\leq1.965$, we rebuild $H(z)$ and the distance modulus for 32 data points using a $\chi^{2}$ minimization strategy. Additionally, we examine $Pantheon + SHOES$ compilation data, which included 1701 SNIa apparent magnitude measurements. Further the \textit{MCMC} analysis has been performed to obtained the best fit values for the model parameters and the EoS parameter. The error bar plots show that the curve for the model and $\Lambda$CDM passing through the range obtained in the graph from the data sets. All the obtained results for the parameters are listed in Table-\ref{table:I}. Further using the $\textit{BAO}$ data set. we measure the expansion of $H(z)$. The constraints on the geometrical parameters are also obtained with the use of data set in order to obtain the accelerating behaviour. The $Om(z)$ diagnostic analysis has been performed, which provides a null test to the $\Lambda$CDM model. 
\par The cosmological model undergoes transition from deceleration to acceleration phase at a transition redshift $z_{t} \approx 0.75$. A deceleration parameter $q_{0} \approx -0.61$ is derived in the model for the transiting Universe at the present epoch. The EoS parameter provides value $-1.16\pm0.17$ and $-1.15\pm0.17$ respectively from the $Hubble$ and $Pantheon + SHOES$ data sets, which shows the phantom behaviour of the model.  We calculated the age of the Universe using the restricted value of the Hubble parameter.  Our derived models were tested against the concordance $\Lambda CDM$ model by recreating the cosmographic parameters and the $Om(z)$ parameter. There may be interactions between dark energy and dark matter components if the model diverges from the $\Lambda CDM$ model. At least at this epoch, the behaviour of the $Om(z)$ parameter in our models may favour a phantom phase.

\section*{Acknowledgement} BM acknowledges the support of IUCAA, Pune (India) through the visiting associateship program. The authors are thankful to Dr. Sunny Vagnozzi,  University of Trento, Italy for his valuable suggestions during the revision of the paper. Nevertheless the authors are thankful to the anonymous reviewers for their comments and suggestions to improve the quality of the paper.
\begin{widetext}
\begin{center}
\begin{table}[H]
\caption{The observational data sets that was used in \cite{Moresco2022}.} 
\centering 
\begin{tabular}{c c c c c | c c c c c} 
\hline\hline 
No. & Redshift & H(z) & $\sigma_{H}$ & Ref. & No. & Redshift & H(z) & $\sigma_{H}$ & Ref.\\ [0.5ex] 
\hline 
1.  & 0.07 & 69.0 & 19.6 & \cite{Zhang2014}      & 17. & 0.4783 & 80.9 & 9.0 &  \cite{Moresco2016} \\ 
2.  & 0.09 & 69.0 & 12.0 & \cite{Simon2005}    & 18. & 0.48 & 97 & 62 &  \cite{Stern2010} \\ 
3.  & 0.12 & 68.6 & 26.2 & \cite{Zhang2014}      & 19. & 0.593 & 104 & 13 & \cite{Moresco2012} \\
4.  & 0.17 & 83 & 8 &  \cite{Simon2005}          & 20. & 0.68 & 92 & 8 & \cite{Moresco2012} \\
5.  & 0.179 & 75.0 & 4.0 & \cite{Moresco2012}    & 21. & 0.75 & 98.8 & 33.6 &  \cite{Borghi2022} \\
6.  & 0.199 & 75.0 & 5.0 &  \cite{Moresco2012}   & 22. & 0.781 & 105 & 12 &  \cite{Moresco2012}\\
7.  & 0.200 & 72.9 & 29.6 &  \cite{Zhang2014}    & 23. & 0.875 & 125 & 17 &  \cite{Moresco2012} \\
8.  & 0.27 & 77 & 14 &  \cite{Simon2005}         & 24. & 0.88 & 90 & 40 &  \cite{Stern2010}  \\
9.  & 0.28 & 88.8 & 36.6 & \cite{Zhang2014}      & 25. & 0.9 & 117 & 23 &  \cite{Simon2005} \\
10. & 0.352 & 83 & 14 &  \cite{Moresco2012}      & 26. & 1.037 & 154 & 20 &  \cite{Moresco2012} \\ 
11. & 0.38 & 83.0 & 13.5 &  \cite{Moresco2016}   & 27. & 1.3 & 168 & 17 &  \cite{Simon2005} \\
12. & 0.4 & 95 & 17 &  \cite{Simon2005}          & 28. & 1.363 & 160 & 33.6 & \cite{Moresco2015} \\
13. & 0.4004 & 77 & 10.2 &  \cite{Moresco2016}   & 29. & 1.43 & 177 & 18 &  \cite{Simon2005} \\
14. & 0.425 & 87.1 & 11.2 &  \cite{Moresco2016}  & 30  & 1.53 & 140 & 14 &  \cite{Simon2005} \\
15. & 0.445 & 92.8 & 12.9 &  \cite{Moresco2016}  & 31. & 1.75 & 202 & 40 &  \cite{Simon2005} \\
16. & 0.47 & 89 & 49.6 & \cite{Ratsimbazafy2017} & 32. & 1.965 & 186.5 & 50.4 & \cite{Moresco2015} \\ 

\hline 
\end{tabular}
\label{table:IV} 
\end{table}
\end{center}
\end{widetext}

  \end{document}